\documentclass[letter]{jpsj2} 
\title{Sb-NQR probe for multipole degree of freedom in the first Pr-based heavy-fermion superconductor PrOs$_{4}$Sb$_{12}$}

\author{ Hideki \textsc{Tou}$^{1}$\thanks{E-mail address:tou@hiroshima-u.ac.jp}, Masahiro \textsc{Doi},$^{1}$ Masafumi \textsc{Sera},$^{1}$ Mamoru \textsc{Yogi},$^{2}$\thanks{Present address:Faculty of Science, University of the Ryukyus, Okinawa 903-0213 }   Hitoshi Sugawara,$^{3}$  Ryousuke Shiina,$^{4}$  and Hideyuki Sato$^{4}$ }

\inst{$^1$ Department of Quantum Matter, AdSM, Hiroshima University, Higashi-Hiroshima, 739-8530 \\
$^2$Department of Materials Science and Technology, Graduate School of Engineering Science, Osaka University, Toyonaka, Osaka 560-8531\\
$^3$Faculty of Integrated Arts and Sciences, The University of Tokushima, Tokushima 770-8502 \\
$^4$Department of Physics, Tokyo Metropolitan University, Hachi-oji 192-0397}

\abst{We report $^{121,123}$Sb nuclear quadrupole resonance (NQR) measurements in the filled-skutterudite superconductor PrOs$_4$Sb$_{12}$ in the temperature range of 0.05-30 K. The electric field gradients (EFG), $V_{zz}$ and $V_{xx}-V_{yy}$,  at the Sb site exhibit unusual temperature dependence below 30 K. To explain these features, we discuss the coupling between the Sb nuclear quadrupole moment and Pr $4f^2$-derived multipole moments. The observed $T$ dependence of EFG is well explained by the CEF quasi-quartet consisted of $\Gamma_1$ singlet and $\Gamma_4^{(2)}$ triplet states in the cubic point group $T_h$. These results, in turn, are indicative of the importance of the coupling between the $^{121,123}$Sb quadrupole moments and the hexadecapole moment caused by the quasi-quartet. 
}

\kword{NQR, EFG, filled-skutterudite, multipole moment, crystal electric field, PrOs$_4$Sb$_{12}$}

\begin{document}
\maketitle


Much interest is drawn to a rich variety of the unconventional nature of the superconducting state in the Pr-based heavy-fermion superconductor PrOs$_4$Sb$_{12}$ with $T_c=1.85$ K \cite{Bauer}. An intriguing feature of the fundamental physics of this compound is strong electron correlations attributed to the Pr $4f^2$ electrons, which is evidenced by both the large electronic specific heat coefficient $\gamma=350\sim700$ mJ/mol K$^2$ \cite{Bauer,Maple} and enhanced cyclotron-effective masses. \cite{Sugawara}  Since a large specific heat jump is observed at $T_c$, the heavy quasiparticles themselves are expected to take part in the formation of the cooper pair. 

Despite such a heavy-fermion character, the magnetic properties of this material have been well explained by a crystal electric field (CEF) model for single-ion Pr 4$f^2$ states. The CEF level scheme in the $T_h$ point group has been established as follows: \cite{Kohgi,Tayama,Goremychkin} a nonmagnetic $\Gamma_1$ singlet ground state (0K), a $\Gamma_4^{(2)}$ triplet first excited state (8K), $\Gamma_4^{(1)}$ triplet (135K) and $\Gamma_{23}$ doublet (205K) higher energy excited states. Recent experiments have shown that this low-lying singlet-triplet scheme leads to a field induced antiferro-quadrupole ordered state at high fields after the superconductivity disappears. \cite{Aoki,Kohgi,Tayama}  Thus, the most foundamental questions for PrOs$_4$Sb$_{12}$ are the properties of these two phases and their mutual relations. 
  
It has been believed that the low-lying  $\Gamma_1$ singlet and $\Gamma_4^{(2)}$ triplet states are the key for controlling the exotic ground state properties. From a group-theoretical perspective, Shiina has discussed the roles of multipole degree of freedom arising from the singlet-triplet states which are regarded as a pseudo-quartet; this quartet causes the field induced ordered state. \cite{Shiina2} From different viewpoints, Otsuki and co-workers have investigated the origin of the CEF splitting by taking account of $p-f$ hybridizations;\cite{Otsuki} they also pointed out the low-lying singlet-triplet states are regarded as a pseudo-quartet leading to multipole orders. Furthermore, Matsumoto and Koga have argued that the pseudo-quartet state is responsible for the appearance of the exotic superconductivity.\cite{Matsumoto} In this context, it is essential to unravel the relationship between the multipole degree of freedom caused by the pseudo-quartet and the unconventional superconductivity in PrOs$_4$Sb$_{12}$. 

To probe the effect of multipole interactions of $f$ electrons, nuclear magnetic resonance (NMR) and nuclear quadrupole resonance (NQR) are effective tools, since nuclear moments of ligand atoms around a rare-earth site can couple with multipole moments of $f$ electrons with the same time-reversal.\cite{Sakai0,Shiina,Sakai} In practice, such a coupling have been discussed in several $f$ electron systems CeB$_6$,\cite{Sakai} NpO$_2$,\cite{Sakai} and URu$_2$Si$_2$.\cite{Takagi} For PrOs$_4$Sb$_{12}$, an importance of a coupling between Sb nuclear quadrupole moments and Pr $4f$ multipole moments has already been pointed out in an Sb NQR study based on the fact that the  temperature ($T$) dependences of the electric field gradient (EFG) and nuclear spin-lattice relaxation rate ($1/T_1$) are anomalous. \cite{Kotegawa}

In this paper, we report a detailed study of $T$ dependence of the EFG at the Sb site in PrOs$_4$Sb$_{12}$ using the Sb NQR. The Sb NQR measurements have been extended to all the transitions for $^{121}$Sb and $^{123}$Sb NQR spectra and to the $T$ range of 0.05-30 K. We have found that the anomalous $T$ dependence of EFG parameters can be explained by taking account of a singlet-triplet CEF state forming a quasi-quartet; the coupling between $^{121,123}$Sb nuclear quadrupole moments and the $4f^2$-derived hexadecapole moment caused by the quasi-quartet is important at low temperatures below 15 K. \cite{Shiina2}

Single crystals of PrOs$_4$Sb$_{12}$ were grown by an Sb-self flux method reported by elsewhere \cite{Sugawara}. $T_c$ of the present sample is 1.83 K. The residual resistivity was obtained to be $\rho_0< 7 \mu\Omega$cm, and the observation of the de Haas-van Alphen (dHvA) signal provides the mean free path of $l\simeq 2000$ {\AA} \cite{Sugawara}, indicating the high quality of the sample. Some of the single crystals were crushed into grains/powder to improve the NQR detection sensitivity. In addition, two single crystals were also used for the present study; sample No.1 has dimensions of $L([110])\times W([1\bar{1}0])\times H([001])\sim 3.2\times3.2\times2.4$ mm$^3$ with a trigonal shape; sample No.2 has $L([100])\times W([010])\times H([001])\sim 1\times1\times3$ mm$^3$ with a rectangular shape. $^{121}$Sb ($I=5/2$) and $^{123}$Sb ($I=7/2$) NQR measurements were carried out on a crushed-grain sample and single crystalline samples using a conventional pulsed spectrometer. The Sb NQR spectra were obtained by tracing the spin-echo intensity as a function of the frequency. The NQR measurements in the $T$ range of 0.4-30 K were carried out at Hiroshima University, while those in the $T$ range of 0.05-1.5 K were done at Osaka University. As long as we use the crushed-grain sample, the NQR linewidth is not different from that of the single crystals; while it becomes approximately double when we use a fine powder.

Figure 1 shows the $^{121,123}$Sb-NQR spectra for two isotopes in the paramagnetic state at 4.2 K. The observed $^{121}$Sb NQR spectrum consists of two narrow lines around the resonance frequencies of $^{121}\nu_1=53.309$ MHz and $^{121}\nu_2=85.0930$ MHz, while the $^{123}$Sb NQR spectrum consists of three lines around $^{123}\nu_1=42.161$ MHz, $^{123}\nu_2=49.910$ MHz and $^{123}\nu_3=78.704$ MHz. The obtained values are consistent with the previous report. \cite{Kotegawa} Here it should be noted that the large value of an asymmetry parameter $\eta\approx 0.45$ \cite{Kotegawa} leads to mixing of pure eigenstates $\Psi_m$ for $\eta=0$ differing by $\Delta m=2$. Here, $m$ represents the magnetic quantum number with respect to the $z$ axis of the principal system. That is, typical eigenstates are given by a linear combination of $\Psi_m$, $e.g.$ $\Phi_m=\sum_m C_m\Psi_m$. For those reasons, each NQR peak frequency is not equivalent to the integral multiplication of the nuclear quadrupole frequency $\nu_Q$

Figure 2 depicts the $T$ dependence of the resonance frequencies $^{123}\nu_{i}$ ($i=1,2,3$). Both  $^{123}\nu_2(T)$ and $^{123}\nu_3(T)$ increase with decreasing $T$, being consistent with the previous report \cite{Kotegawa}. On the other hand, it is noteworthy that $^{123}\nu_1(T)$ exhibits opposite $T$ dependence.  $^{121}\nu_{i}$ ($i=1,2$) are also dependent of $T$ (not shown). These features indicate that  $\eta$ as well as $\nu_Q$ changes with $T$.

In order to deduce more accurate NQR parameters, the resonance frequencies for $^{121,123}$Sb nuclei of  PrOs$_4$Sb$_{12}$ were calculated by exact diagonalization of the nuclear spin Hamiltonian matrix of $^{121,123}$Sb as,\cite{tou0}
\begin{equation}
{\cal H}_Q= \frac{1}{6}h\nu_Q\left[3I_z^2-I(I+1)+\frac{\eta}{2}(I_+^2+I_-^2)\right] ,
\end{equation}
where  $\nu_Q\equiv \frac{3e^2qQ}{2I(2I-1)h}$ and $\eta\equiv \frac{V_{xx}-V_{yy}}{V_{zz}}$. $e$ is the electron charge, and $Q$ is the nuclear quadrupole moment, $V_{\alpha\alpha}=\frac{\partial^2V}{\partial \alpha^2}$ is the EFG for $\alpha(=x, y, z)$ direction, and $eq=V_{zz}$ is the maximum EFG, where the principal axes of the EFG  are defined according to the previous paper.\cite{tou} The calculations yield $^{121}\nu_Q=44.177$ MHz, $^{123}\nu_Q=26.817$ MHz, and $\eta=0.4586 $ at $T=4.2$ K. To obtain the $T$ dependence of $\eta$ and $\nu_Q$, we calculated the resonance frequencies for $^{121,123}$Sb at various temperatures using Eq.(1). The $T$ dependence of $^{121}\nu_Q(T)$ and $\eta(T)$ is shown in Fig. 3. The remarkable feature in the NQR parameters is the strong $T$ dependence of both $^{121}\nu_Q(T)$ and $\eta(T)$; it is the first case in our knowledge.

The NQR frequency $\nu_Q(T)$ in metals without $f$ electrons shows normally a monotonic increase with decreasing $T$. In such a case, $\nu_Q(T)$ is often discussed in terms of an empirical relation of $\nu_Q(T)=\nu_Q(0)(1-\alpha T^{3/2})$ with $\alpha>0$.\cite{Takagi2}  Since this relation shows a monotonic increase with decreasing $T$ from high $T$ as high as a room temperature, it cannot explain the abrupt increase of $\nu_Q(T)$ below 10 K in PrOs$_4$Sb$_{12}$. Alternatively, the $T$ dependence of both $\nu_Q(T)$ and $\eta(T)$ in PrOs$_{4}$Sb$_{12}$ could be attributed to the coupling between the $^{121,123}$Sb nuclear quadrupole moments and Pr $4f^2$-derived multipole moments. \cite{Sakai,Takagi}

To discuss the coupling of the $^{121,123}$Sb nuclear quadrupole moments and Pr $4f^2$-derived multipole moments, we need a CEF Hamiltonian for the cubic point group $T_h$ in PrOs$_4$Sb$_{12}$, which is given as ${\cal H}_{CEF}=A_4(O_4^0+5O_4^4)+A_6(O_6^0-21O_6^4)+A_6'(O_6^2-O_6^6)$, where $A_4$, $A_6$, and $A_6'$ are the CEF parameters and $O_l^m$ the Stevens operators associated with the multipole moments. \cite{Takegahara,stevens}

In PrOs$_{4}$Sb$_{12}$, the level splitting between the singlet $\Gamma_1$ state and triplet $\Gamma_4^{(2)}$ 
first excited states is estimated as $\Delta \sim 8$K.\cite{Kohgi} Since the excitation energies to higher levels are considered as more than 100K, the low $T$ properties of PrOs$_{4}$Sb$_{12}$ would be governed by the $\Gamma_1$-$\Gamma_4^{(2)}$ pseudo-quartet state.\cite{Shiina2,Otsuki} Such a quartet state has fifteen multipole moments consisting of three dipoles, five quadrupoles, three octupoles, and four hexadecapoles (See Table II in Ref.(8)). Those multipoles of $f$ electrons can couple with $^{121,123}$Sb nuclear moments with the same time reversal. In the paramagnetic phase, only a hexadecapole $H^0$ that belongs to a $\Gamma_1$ representation has a finite thermal average, so that it can contribute to a shift of the asymmetry parameter as well as the NQR frequency. In fact, it is not difficult to make up an invariant hyperfine coupling between $H^0$ and the $^{121,123}$Sb nuclear quadrupole moments on the basis of the Pr-Sb local bond symmetry. \cite{Shiina,Shiina2,Sakai,Takagi}

Let us examine the effect of a hexadecapole $H^0$ in detail. The symmetry analysis allows us to 
write the EFG parameters with the following expressions, 
\begin{eqnarray}
&&V_{zz}(T)=V_{zz}(0)+a\langle H^0 \rangle, \nonumber \\ 
&&V_{xx-yy}(T)=V_{xx-yy}(0)+b\langle H^0 \rangle,
\end{eqnarray}
making use of $V_{zz}(0)$, $V_{xx-yy}(0)$, $a$, and $b$ as fitting parameters. Here $V_{xx-yy}$ is defined by $V_{xx-yy} \equiv V_{xx}-V_{yy}$.  Within the pseudo-quartet model, the thermal average of $H^0$ is given by 
a simple distribution between the ground singlet and the excited triplet as 
\begin{eqnarray}
\langle H^0 \rangle&\propto &\langle O_4^0+5O_4^4 \rangle, \nonumber \\ 
&\propto&\frac{3e^{-\Delta/k_BT}}{1+3e^{-\Delta/k_BT}},
\end{eqnarray}
where $k_B$ is the Boltzmann factor and $\Delta$ the level splitting ($\Delta=8$K). 
In the following we shall compare this formula with the observed NQR results.

Figure 4 represents the $T$ dependence of $V_{zz}$ and $V_{xx-yy}$, where we use the values of Sb quadrupole moments, $^{121}Q=-0.597\times 10^{-28}$ m$^2$, and $^{123}Q=-0.762\times10^{-28}$ (m$^2$). \cite{tou} The solid lines are calculated curves using Eqs.(2) and (3) below 15 K; the fitting parameters are obtained as $a=4.76\times 10^{19}$, $b=-5.26\times 10^{19}$, $V_{zz}(0)=-2.042\times10^{22}$, $V_{xx-yy}(0)=-9.34\times10^{21}$. 

Clearly, the agreements between data and  $V_{zz}(T)$ and $V_{xx-yy}(T)$ are very good below 10 K. Our data, therefore, provide evidence that the local electronic state at the Sb site at low temperatures is dominated by the singlet-triplet CEF states. Alternatively, the microscopic origin of the $T$ dependence of the EFG parameters can be explained by taking account of the coupling of the $^{121,123}$Sb nuclear quadrupole moments with the hexadecapole moment under the same time reversal symmetry. 

The deviations above 15 K cannot be explained by the inclusion of the triplet $\Gamma_4^{(1)}$ second excited state (135 K). Recently Goto and co-workers have detected a thermally activated rattling motion due to the off-center Pr-ion by means of ultra sonic measurements, where the rattling motion is not random, but has $\Gamma_{23}$ symmetry.\cite{Goto} Such an anisotropic motion might deform the $4f$ electron charge distrbution, hence it would cause the decoupling between the Pr 4f hexadecapole moment and Sb nuclear quadrupole moment. For that reason, we assume the fitting parameters $a$ and $b$ in Eq.(2) are replaced as $a\{1-\alpha/(1+\omega^2\tau^2)\}$ and $b\{1-\beta/(1+\omega^2\tau^2)\}$, respectively, by introducing a decoupling term with the Debye type dispersion.\cite{Goto}  Here $\alpha$ and $\beta$ are decoupling parameters, $\omega$ is the angular frequencies, $\tau$ is the correlation time $\tau=\tau_0\exp(E/k_BT)$ with an attempt time $\tau_0$ and an activation energy $E$ for a motion. The calculated curves (dashed lines) in the Fig. 4 reproduces well the experimental data; the fitting parameters for the rattling motion are obtained as $\alpha=0.15$, $\beta=0.4$, and $E=80$ K, assuming $\tau_0=8.8\times10^{-11}$ s \cite{Goto} and $\omega/2\pi=42$ MHz. Good agreements suggest that the rattling motion of Pr-ion is a possible origin for the deviations above 15 K. Since this thermally activated rattling motion quenches at low temperature below 15 K, it is evident that the low temperature properties of this compound is governed by the multipole degree of freedom. Here, it is worth emphasizing that the present Sb NQR results give evidence for the singlet-triplet CEF level scheme reported previously; \cite{Kohgi,Tayama,Goremychkin,Aoki} the other CEF ground states cannot reproduce the present Sb NQR results.

The present Sb NQR results also give evidence for the multipole degree of freedom caused by the CEF quasi-quartet state from microscopic viewpoints. The multipole degree of freedom possibly leads to the multichannel Kondo effect that conduction electrons interact with $f$-electrons through orbital channels such as multipoles.\cite{Cox} Such a multichannel Kondo effect in the periodic lattice, i.e., periodic multichannel Kondo effect, might cause the  exotic ground states controlled by magnetic fields in PrOs$_4$Sb$_{12}$, i.e., the unconventional heavy-fermion superconductivity and the field induced order.  In this context, multipole fluctuations, e.g., quadrupole fluctuations, arising from the multipole degree of freedom are a candidate for the medium of the Cooper pairing in PrOs$_4$Sb$_{12}$. Actually, Kotegawa et. al. have pointed out that Pr $4f^2$-derived quadrupole fluctuations are a key to understand the appearance of the unconventional superconductivity in  PrOs$_4$Sb$_{12}$ from the unusual $T$ dependence of the NQR relaxation rate in the paramagnetic phase.\cite{Kotegawa} To prove the multipole fluctuations, further measurements are required.  

In summary, we have performed $^{121,123}$Sb NQR  measurements in the filled-skutterudite superconductor PrOs$_4$Sb$_{12}$ in the $T$ range of 0.05-30 K. The EFG parameters, $V_{zz}$ and $V_{xx}-V_{yy}$, at the Sb site exhibit unusual $T$ dependence below 30 K. To explain this feature, we have discussed the coupling between the Sb nuclear quadrupole moment and Pr $4f^2$-derived multipole moments; the $T$ dependence of the  EFG parameters can be explained by the coupling of the Sb nuclear quadrupole moment and the Pr $4f^2$-derived hexadecapole moment which has a finite thermal average in the paramagnetic state. In the present case, it has turned out to be identical to the thermal average for the $\Gamma_1$-$\Gamma_4^{(2)}$ CEF pseudo-quartet state. The evidence for the multipole degree of freedom provides a crucial clue to clarify the intimate relationship between the field induced antiferro quadrupole order and the unconventional superconductivity in PrOs$_4$Sb$_{12}$.  Further experimental and theoretical efforts are needed to unravel the microscopic origin of the multipole order and unconventional superconductivity in PrOs$_{4}$Sb$_{12}$. 

\section*{Acknowledgements}
We gratefully acknowledge Y. Kitaoka, G.-q. Zheng, H. Kotegawa, Y. Aoki, H. Harima, S. Takagi, and O. Sakai for valuable comments and discussions. This work was supported by Grants-in-Aid for COE Research (No. 13CE2002) and for Scientific Research Priority Area ``Skutterudite'' (Nos.16037211 and 15072206) of the Ministry of education, Culture, Sports, Science and Technology Japan.

\begin{figure}[h]
\includegraphics[width=0.9\linewidth]{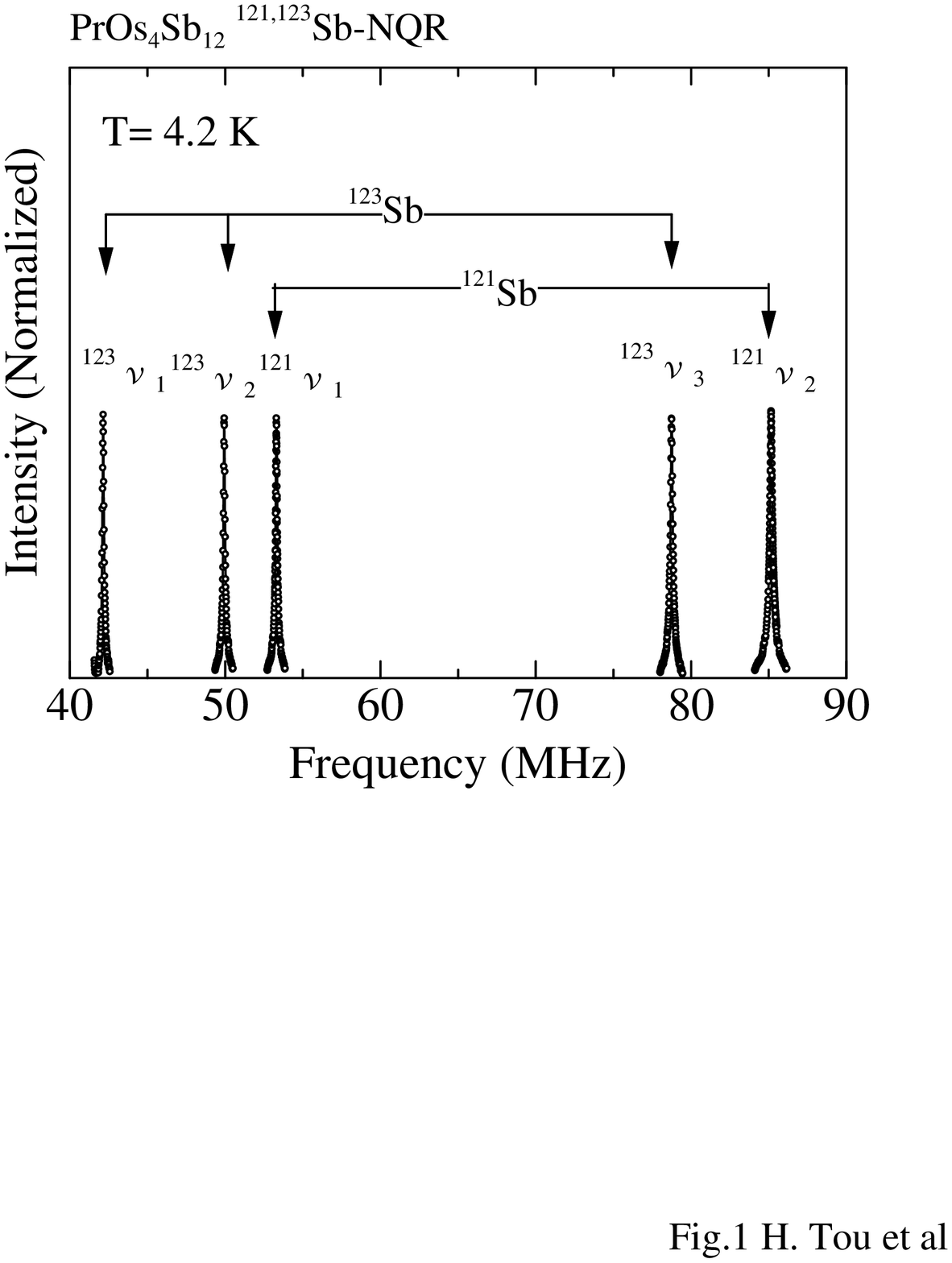}
\caption{$^{121}$Sb and $^{123}$Sb-NQR spectra  in PrOs$_{4}$Sb$_{12}$ at $T=4.2$ K. }
\end{figure}

\begin{figure}[h]
\includegraphics[width=0.9\linewidth]{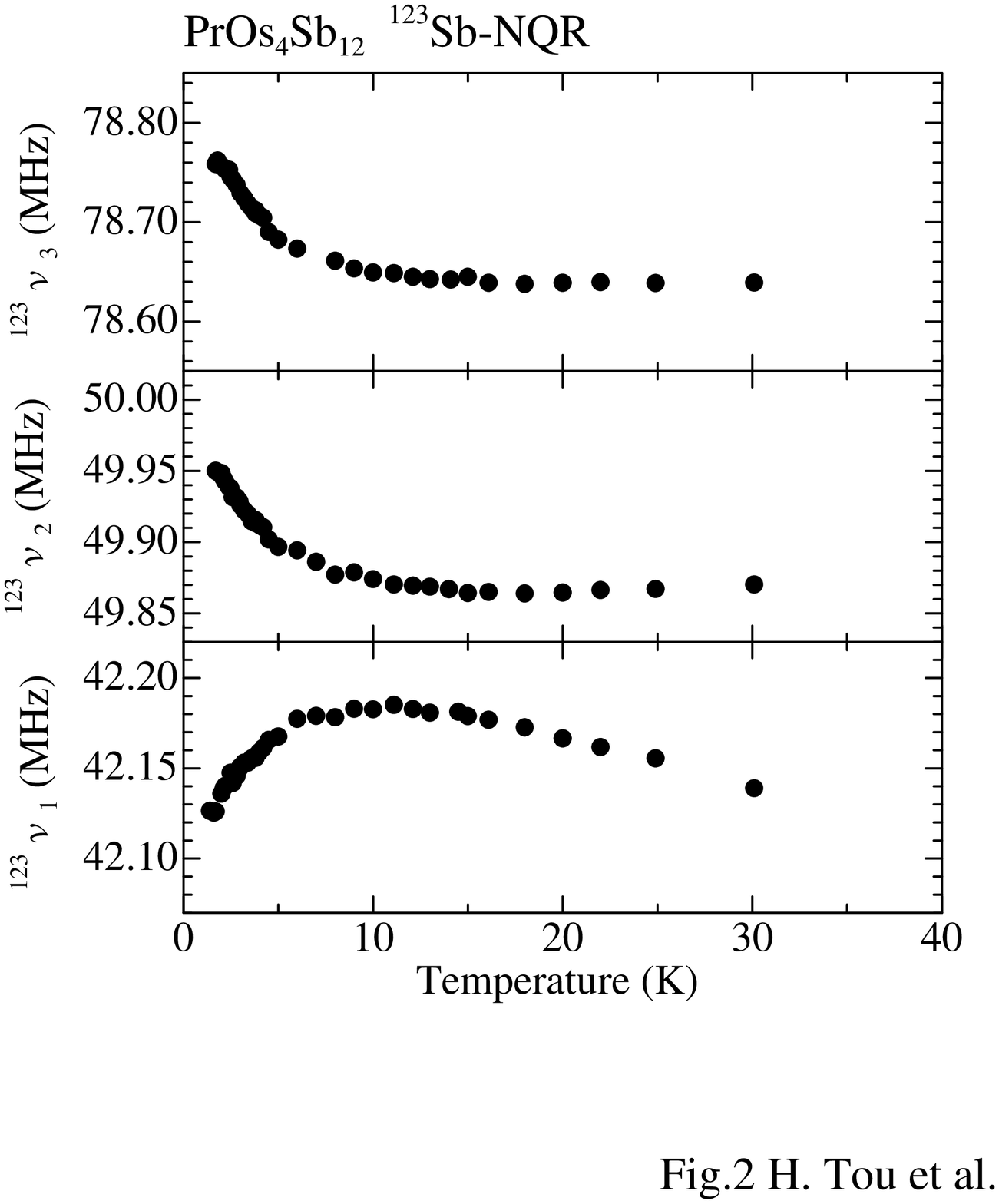}
\caption{ Temperature dependence of the $^{123}$Sb-NQR resonance peaks in PrOs$_{4}$Sb$_{12}$.}
\end{figure}

\begin{figure}[h]
\includegraphics[width=0.9\linewidth]{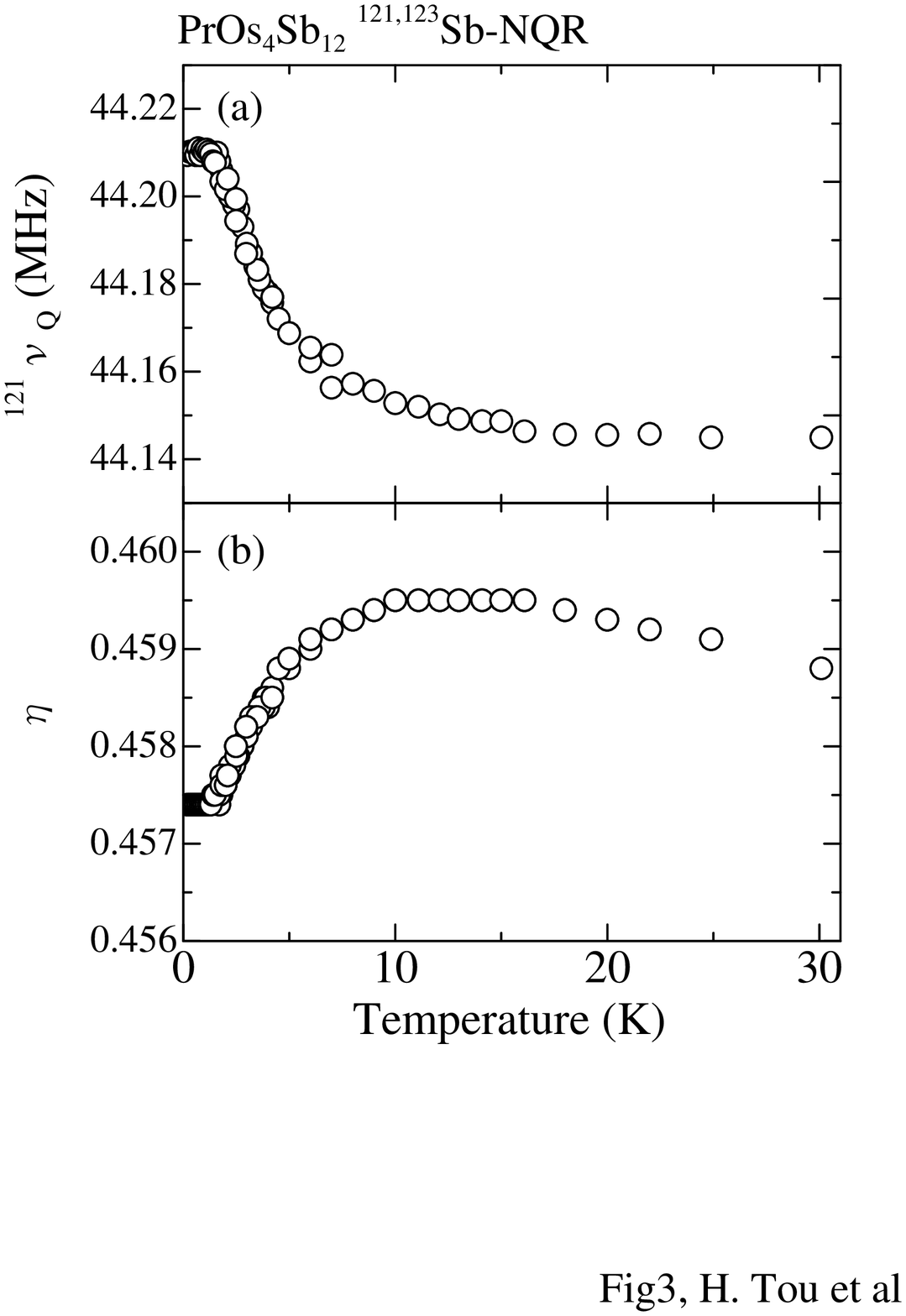}
\caption{Temperature dependence of (a) $^{121}$Sb-NQR frequency $^{121}\nu_Q$ and (b) asymmetry parameter $\eta$ in PrOs$_{4}$Sb$_{12}$. }
\end{figure}

\begin{figure}[h]
\begin{center}
\includegraphics[width=0.9\linewidth]{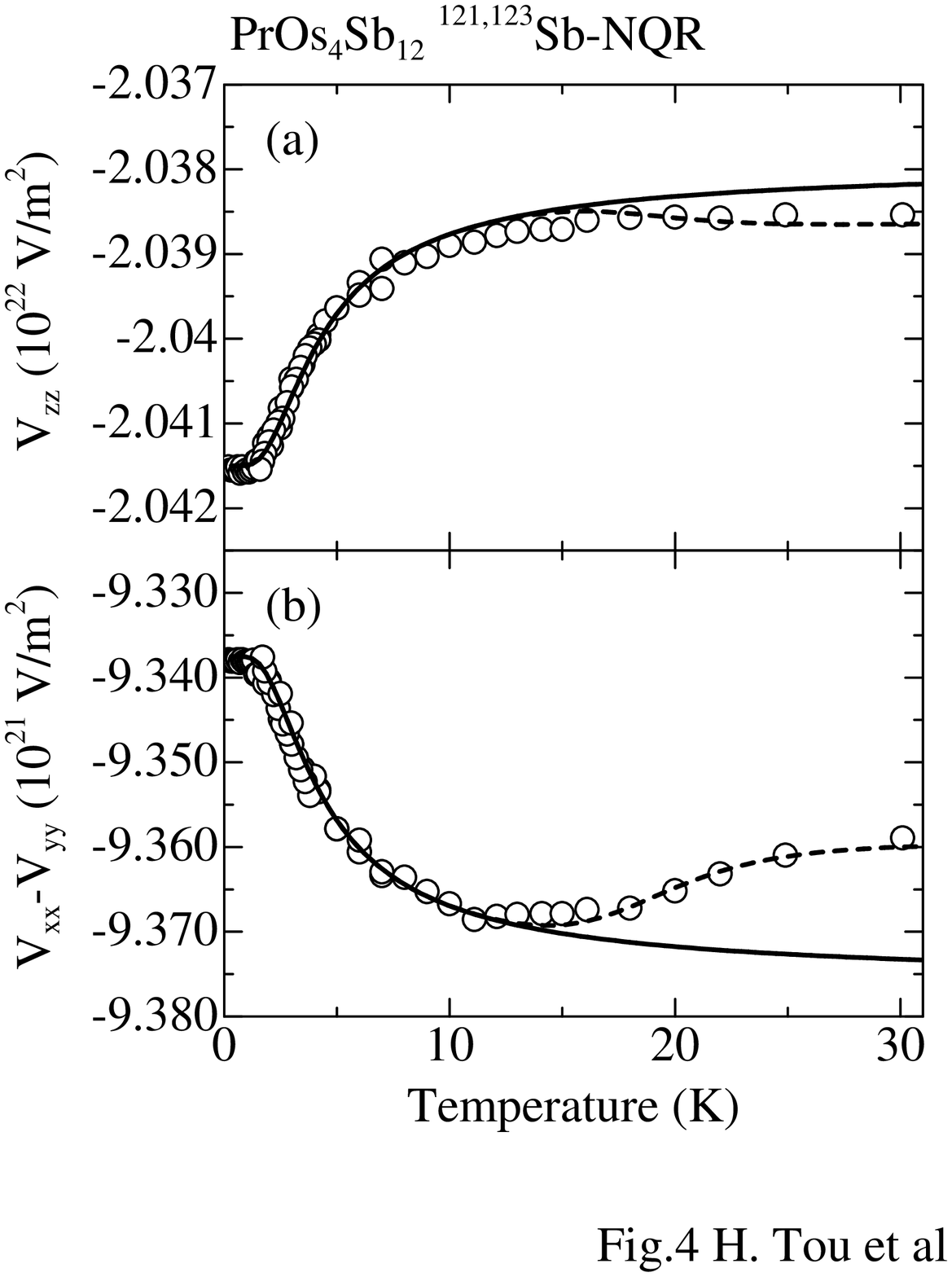}
\end{center}
\caption{Temperature dependence of the EFG parameters, (a) $V_{zz}$ and (b) $V_{xx}-V_{yy}$ in PrOs$_{4}$Sb$_{12}$. The solid lines are fits by Eqs. (3) and (4). The dashed curves represent fits with the contribution from the rattling due to Pr-ion off-center motion. (See text)}
\end{figure}

\end{document}